\documentclass[twocolumn,prl,showpacs,floatfix]{revtex4}

\usepackage{subfigure}
\usepackage{graphicx}

\begin{document}


\title{Bose-Einstein Condensation of Dark Matter Axions}

\author{P. Sikivie and Q. Yang}

\affiliation{Department of Physics, University of Florida, 
Gainesville, FL 32611, USA}

\begin{abstract}
We show that cold dark matter axions thermalize and form a Bose-Einstein
condensate.  We obtain the axion state in a homogeneous and isotropic 
universe, and derive the equations governing small axion perturbations.  
Because they form a BEC, axions differ from ordinary cold dark matter in 
the non-linear regime of structure formation and upon entering the horizon.  
Axion BEC provides a mechanism for the production of net overall rotation 
in dark matter halos, and for the alignment of cosmic microwave anisotropy
multipoles.

\end{abstract}

\maketitle

Several authors have proposed that the dark matter of the universe is 
a Bose-Einstein condensate (BEC) \cite{dmBEC1,dmBEC2}.  The axion is 
sometimes mentioned in this context.  Indeed the axion is a boson and 
a cold dark matter candidate, and cold dark matter axions are known to 
have huge phase space density.  But, as far as we are aware, it has 
never been shown that dark matter axions form a BEC.  Their phase space 
density is certainly large enough but they will only form a BEC if they 
reach thermal equilibrium.  This may see unlikely because the axion is 
very weakly coupled.  Below we find that dark matter axions do form 
a BEC, marginally because of their self interactions but certainly as 
a result of their gravitational interactions.  No special assumptions 
are required.

Shortly after the Standard Model of elementary particles was established,
the axion was postulated \cite{axion} to explain why the strong 
interactions conserve the discrete symmetries P and CP.  For the 
purposes of this paper the action density for the axion field 
$\varphi(x)$ may be taken to be 
\begin{equation}
{\cal L}_a = - {1 \over 2} \partial_\mu \varphi \partial^\mu \varphi
- {1 \over 2} m^2 \varphi^2 + {\lambda \over 4!} \varphi^4 - ... 
\label{lag}
\end{equation}
where $m$ is the axion mass.  The self-coupling strength is 
\begin{equation}
\lambda = {m^2 \over f^2}~{m_d^3 + m_u^3 \over (m_d + m_u)^3}
\simeq 0.35~{m^2 \over f^2}
\label{self}
\end{equation}
in terms of the axion decay constant $f$ and the masses $m_u$ and
$m_d$ of the up and down quarks.  In Eq.~(\ref{lag}), the dots 
represent higher order axion self-interactions and interactions 
of the axion with other particles.  All axion couplings and the 
axion mass
\begin{equation}
m \simeq 6 \cdot 10^{-6}~{\rm eV}~{10^{12}~{\rm GeV} \over f}
\label{mass}
\end{equation}
are inversely proportional to $f$. $f$ was first thought to be of order 
the electroweak scale, but its value is in fact arbitrary \cite{invis}.
However, the combined limits from unsuccessful searches in particle 
and nuclear physics experiments and from stellar evolution require 
$f \gtrsim 3 \cdot 10^9$ GeV \cite{axrev}.  

Furthermore, an upper limit $f \lesssim 10^{12}$ GeV is provided by 
cosmology because light axions are abundantly produced during the QCD
phase transition \cite{axdm}.  In spite of their very small mass, these 
axions are a form of cold dark matter.  Indeed, their average momentum 
at the QCD epoch is not of order the temperature (GeV) but of order the 
Hubble expansion rate ($3 \cdot 10^{-9}$ eV) then.  In case inflation 
occurs after the Peccei-Quinn phase transition their average momentum is 
even smaller because the axion field gets homogenized during inflation.  
For a detailed discussion see ref. \cite{axcos}.  In addition to this 
cold axion population, there is a thermal axion population with average 
momentum of order the temperature.  

The non-perturbative QCD effects that give the axion its mass turn on
at a temperature of order 1 GeV.  The critical time, defined by 
$m(t_1) t_1 = 1$, is $t_1 \simeq 2 \cdot 10^{-7}~{\rm sec}~
(f / 10^{12}~{\rm GeV})^{1 \over 3}$.  Cold axions are the quanta 
of oscillation of the axion field that result from the turn on of 
the axion mass.  They have number density 
\begin{equation}
n(t) \sim {4 \cdot 10^{47} \over {\rm cm}^3}~
\left({f \over 10^{12}~{\rm GeV}}\right)^{5 \over 3}
\left({a(t_1) \over a(t)}\right)^3
\label{numden}
\end{equation}
where $a(t)$ is the cosmological scale factor.  Because the axion 
momenta are of order ${1 \over t_1}$ at time $t_1$ and vary with 
time as $a(t)^{-1}$, the velocity dispersion of cold axions is 
\begin{equation}
\delta v (t) \sim {1 \over m t_1}~{a(t_1) \over a(t)}
\label{veldis}
\end{equation} 
{\it if} each axion remains in whatever state it is in, i.e. if axion 
interactions are negligible.  Let us refer to this case as the limit of 
decoupled cold axions.  If decoupled, the average state occupation number 
of cold axions is 
\begin{equation}
{\cal N} \sim~ n~{(2 \pi)^3 \over {4 \pi \over 3} (m \delta v)^3}
\sim 10^{61}~\left({f \over 10^{12}~{\rm GeV}}\right)^{8 \over 3}~~\ .
\label{occnum}
\end{equation}
Clearly, the effective temperature of cold axions is much smaller than 
the critical temperature  
\begin{equation}
T_{\rm c} = \left({\pi^2 n \over \zeta(3)}\right)^{1 \over 3}
\simeq 300~{\rm GeV}~\left({f \over 10^{12}~{\rm GeV}}\right)^{5 \over 9}~
{a(t_1) \over a(t)}
\label{Tc}
\end{equation}
for BEC.  Axion number violating processes, such as their decay to two 
photons, occur only on time scales vastly longer than the age of the 
universe.  The only condition for axion BEC that is not manifestly 
satisfied is thermal equilibrium.

Axions are in thermal equilibrium if their relaxation rate $\Gamma$ is 
large compared to the Hubble expansion rate $H(t) = {1 \over 2t}$.  At 
low phase space densities, the relaxation rate is of order the particle 
interaction rate $\Gamma_s = n \sigma \delta v$ where $\sigma$ is the 
scattering cross-section.  The cross-section for 
$\varphi + \varphi \rightarrow \varphi + \varphi$ scattering due to 
axion self interaction is {\it in vacuum} 
\begin{equation}
\sigma_0 = {1 \over 64 \pi} {\lambda^2 \over m^2} \simeq
1.5 \cdot 10^{-105} {\rm cm}^2 \left({m \over 10^{-5}~{\rm eV}}\right)^6~~~\ .
\label{xs0}
\end{equation}
If one substitutes $\sigma_0$ for $\sigma$, $\Gamma_s$ is found much 
smaller than the Hubble rate, by many orders of magnitude.  However, 
in the cold axion fluid background, the scattering rate is enhanced by 
the average quantum state occupation number of both final state axions, 
$\sigma \sim \sigma_0 {\cal N}^2$, because energy conservation forces 
the final state axions to be in highly occupied states if the initial 
axions are in highly occupied states.  In that case, the relaxation rate
is multiplied by {\it one} factor of ${\cal N}$ \cite{ST}
\begin{equation}
\Gamma \sim n~\sigma_0~\delta v~{\cal N}~~~\ .
\label{rate}
\end{equation}
Combining Eqs.~(\ref{numden}-\ref{occnum},\ref{xs0}), one finds 
$\Gamma(t_1)/H(t_1) \sim {\cal O}(1)$, suggesting that cold axions
thermalize at time $t_1$ through their self interactions, but only 
barely so.  

It may seem surprising that the huge and tiny factors on the RHS of 
Eq.~(\ref{rate}) cancel each other.  In fact the cancellation is not 
an accident.  Consider a generic axion-like particle (ALP) whose mass 
$m$ and decay constant $f$ are unrelated to each other.  Its self 
interaction coupling strength $\lambda \sim {m^2 \over f^2}$.  Cold 
ALPs appear at a time $t_1 \sim {1 \over m}$ with number density 
$n(t_1) \sim f^2 m$, and velocity dispersion $\delta v (t_1) \sim 1$.  
Substituting these estimates in Eqs. (\ref{occnum}), (\ref{xs0}) and 
(\ref{rate}), one finds that the thermalization rate is of order the 
Hubble rate at $t_1$, for all $f$ and $m$.  

A critical aspect of axion BEC phenomenology is whether the BEC 
continues to thermalize after it has formed.  Axion BEC means 
that (almost) all axions go to one state.  However, only if the
BEC continually rethermalizes does the axion state track the 
lowest energy state.

The particle kinetic equations that yield Eq.~(\ref{rate}) are 
valid only when the energy dispersion ${1 \over 2} m (\delta v)^2$ 
is larger than the thermalization rate \cite{ST}.  After $t_1$ this 
condition is no longer satisfied.  One enters then a regime where 
the relaxation rate due to self interactions is of order 
\begin{equation}
\Gamma_\lambda  \sim \lambda~n~m^{-2}~~\ .
\label{rate2}
\end{equation} 
$\Gamma_\lambda(t)/H(t)$ is of order one at time $t_1$ but 
decreases as $t~a(t)^{-3}$ afterwards.  Hence, self interactions 
are insufficient to cause axion BEC to rethermalize after $t_1$ 
even if they cause axion BEC at $t_1$.  However gravitational 
interactions, which are long range, come in to play.  The 
relaxation rate due to gravitational interactions is of order
\begin{equation}
\Gamma_{\rm g} \sim G~n~m^2~\ell^2
\label{rate3}
\end{equation}
where $\ell \sim (m \delta v)^{-1}$ is the correlation length.  
$\Gamma_{\rm g}(t)/H(t)$ is of order 
$4 \cdot 10^{-8}(f/10^{12}~{\rm GeV})^{2 \over 3}$ 
at time $t_1$ but grows as $t a^{-1}(t) \propto a(t)$.  Thus 
gravitational interactions cause the axions to thermalize and 
form a BEC when the photon temperature is of order 
100 eV~$(f/10^{12}~{\rm GeV})^{1 \over 2}$.

The process of axion Bose-Einstein condensation is constrained by 
causality.  We expect overlapping condensate patches with typical 
size of order the horizon.  As time goes on, say from $t$ to $2t$, 
the axions in $t$-size condensate patches rethermalize into $2t$-size 
patches.  The correlation length is then of order the horizon at 
all times, implying $\delta v \sim {1 \over m t}$ instead of 
Eq.~(\ref{veldis}), and $\Gamma_{\rm g}/H \propto t^3 a^{-3}(t)$ 
after the BEC has formed.  Therefore gravitational interactions 
rethermalize the axion BEC on ever shorter time scales compared 
to the age of the universe.

We now consider what implications axion BEC has for observation.  
The axion field may be expanded in modes 
labeled $\vec\alpha$:
\begin{equation}
\varphi(x) = \sum_{\vec\alpha}~[a_{\vec\alpha}~\Phi_{\vec\alpha}(x)
~+~a_{\vec\alpha}^\dagger~\Phi_{\vec\alpha}^\star]
\label{modex}
\end{equation}
where the $\Phi_{\vec\alpha}(x)$ are the positive frequency c-number 
solutions of the Heisenberg equation of motion for the axion field 
\begin{equation}
D^\mu D_\mu \varphi(x) = g^{\mu\nu}[\partial_\mu \partial_\nu -
\Gamma_{\mu\nu}^\lambda \partial_\lambda] \varphi(x) = m^2 \varphi(x)~~~\ ,
\label{eom}
\end{equation}
and the $a_{\vec\alpha}$ and $a_{\vec\alpha}^\dagger$ are creation 
and annihilation operators satisfying canonical commutation relations. 
We neglect the self-interaction term
$- {1 \over 6} \lambda \varphi^3$, which would otherwise appear on 
the RHS of Eq.~(\ref{eom}), because it is of order 
${\rho \over f^2} \varphi$, where $\rho$ is the axion density,
and hence smaller by the factor
$\left({a(t_1) \over a(t)}\right)^3 {t \over t_1}$ than the relevant
terms (of order ${m \over t} \varphi$) in that equation.  When the 
self-interactions are included, one finds an instability in the axion 
BEC towards the formation of droplets.  The analog of the sound speed 
\cite{sound} is imaginary in this case because the self interaction is 
attractive.  However,  the rate of droplet formation is negligibly small 
compared to the Hubble rate.  The gravitational forces always dominate 
over the self interactions except briefly after the cold axions first 
appear at time $t_1$

Except for a tiny fraction, all cold axions go to a single state 
which we label $\vec\alpha = 0$.  The corresponding $\Phi_0(x)$ 
is the axion wavefunction.  In the spatially flat, homogeneous 
and isotropic Robertson-Walker space-time,
\begin{equation}
\Phi_0 = {A \over a(t)^{3 \over 2}}~e^{-i m t}
\label{RW}
\end{equation}
where $A$ is a constant.  The state of the axion field is 
$|N> = (1/\sqrt{N!})~(a_0^\dagger)^N |0>$ where $|0>$ is the 
empty state, defined by $a_{\vec\alpha}~|0>$ = 0 for all 
$\vec\alpha$, and $N$ is the number of axions.  The 
expectation value of the stress-energy-momentum 
tensor is
\begin{eqnarray} 
<N|T_{\mu\nu}|N> &=& N
[\partial_\mu \Phi_0^* \partial_\nu \Phi_0 \nonumber\\
+ \partial_\nu \Phi_0^* \partial_\mu \Phi_0 &+& 
g_{\mu\nu} ( - \partial_\lambda \Phi_0^* \partial^\lambda \Phi_0
- m^2 \Phi_0^* \Phi_0)]~~~\ .
\label{Tmunu}
\end{eqnarray}
Again we neglect the self-interaction term.  

Consider first the behavior of axion BEC in a flat Minkowski 
space-time.  Since the axions are non-relativistic, 
$\Phi_0(x) = e^{- i m t} \Psi(x)$ with $\Psi(x)$ slowly 
varying.  Neglecting terms of order ${1 \over m}~\partial_t$ 
compared to terms of order one, Eq.~(\ref{eom}) becomes the 
Schr\"{o}dinger equation:   
\begin{equation}
i \partial_t \Psi = - {\nabla^2 \over 2 m} \Psi~~~~\ .
\label{Schrod}
\end{equation}
It is useful \cite{Pethik} to write the wavefunction as 
\begin{equation}
\Psi(\vec{x}, t) = {1 \over \sqrt{2 m N}} B(\vec{x}, t) 
e^{i \beta(\vec{x}, t)}~~~\ .
\label{dec}
\end{equation}
In terms of $B(\vec{x}, t)$ and $\beta(\vec{x}, t)$ the 
energy and momentum densities are ($j,k = 1,2,3$) 
$T_{00} \equiv \rho = m \left(B(\vec{x}, t)\right)^2$ 
and $T_{0j} \equiv - \rho v_j = 
- \left(B(\vec{x}, t)\right)^2 \partial_j \beta$,
in the non-relativistic limit.  The velocity field is therefore 
$\vec{v}(\vec{x}, t) = {1 \over m} \vec\nabla \beta(\vec{x}, t)$ 
\cite{Pethik}.  Eq.~(\ref{Schrod}) implies the continuity equation 
and the equation of motion
\begin{equation}
\partial_t v^k + v^j \partial_j v^k = - \vec\nabla q 
\label{newf}
\end{equation}
where
\begin{equation}
q(\vec{x}, t) = - {\nabla^2 \sqrt{\rho} \over 2 m^2 \sqrt{\rho}}~~~\ .
\label{q}
\end{equation}
Following the motion, the stress tensor is 
\begin{equation}
T_{jk} = \rho v_j v_k + 
{1 \over 4 m^2}({1 \over \rho} \partial_j \rho \partial_k \rho
- \delta_{jk} \nabla^2 \rho)~~~~\ .
\label{stress}
\end{equation}
For ordinary cold dark matter (CDM) the last terms on the RHS of 
Eqs.~(\ref{newf}) and (\ref{stress}) are absent.

To compare axion BEC with CDM we divide the observations into three
arenas: 1) the behaviour of density perturbations on the scale of the 
horizon, 2) their behaviour during the linear regime of evolution within 
the horizon, and 3) their behaviour during the non-linear regime.  We 
first discuss arena 2 where CDM provides a very successful description.
Neglecting second order terms, the perturbation in the stress tensor 
implied by Eq.~(\ref{stress}) is
\begin{equation} \delta T_{jk} = 
- \delta_{jk} {\rho_0(t) \over 4 m^2} \nabla^2 \delta (\vec{x}, t) 
\label{stresspert} 
\end{equation} 
where $\rho_0(t)$ is the unperturbed axion density and 
$\delta (\vec{x}, t) \equiv {\delta \rho (\vec{x}, t) \over \rho_0(t)}$.  
Because the RHS of Eq.~(\ref{stresspert}) is proportional to the Kronecker 
symbol and the RHS of Eq.~(\ref{newf}) is a gradient, vector and tensor 
perturbations are not affected by the additional forces associated with 
the axion BEC.  Only the scalar perturbations are affected.  The scalar 
perturbations are conveniently described in conformal Newtonian gauge 
\cite{Dodelson} where the metric is
\begin{equation} 
ds^2 = - (1 + 2 \psi(\vec{x}, t)) dt^2 
+ a(t)^2 (1 + 2 \phi(\vec{x}, t)) d \vec{x} \cdot d \vec{x}~\ . 
\label{cN}
\end{equation}
Conservation of energy and momentum in this background implies the first 
order equations 
\begin{eqnarray} 
\partial_t \delta + {1 \over a} \vec\nabla \cdot \vec{v} &=&
- 3 \partial_t \phi + {3 H \over 4 m^2 a^2} \nabla^2 \delta\nonumber\\
\partial_t \vec{v} + H \vec{v} &=& - {1 \over a} \vec\nabla \psi + 
{1 \over 4 m^2 a^3} \vec\nabla~\nabla^2 \delta 
\label{denev} 
\end{eqnarray} 
where $H = {1 \over a} {da \over dt}$.  The equations for CDM are 
recovered by letting $m \rightarrow \infty$.  The RHS of Einstein's 
equations are modified by the addition of $\delta T_{jk}$ to the stress 
tensor, but this modification does not play a role in our discussion 
because it is suppressed, relative to the leading terms, by the factor 
$\left({k_{\rm ph} \over m}\right)^2$, where $k_{\rm ph}$ is the physical 
wavevector of the perturbation.

It is clear from Eqs.~(\ref{denev}) that axion BEC differs from CDM 
on small scales only.  For scales that are well within the horizon 
($k_{\rm ph} >> H$), Eqs.~(\ref{denev}) plus Einstein's equations imply 
\begin{equation}
\partial_t^2 \delta + 2 H \partial_t \delta 
- \left(4 \pi G \rho_0 - {k^4 \over 4 m^2 a^4}\right) \delta = 0
\label{deneq}
\end{equation}
for the Fourier components $\delta(\vec{k}, t)$ of $\delta(\vec{x}, t)$.
$\vec{k} = a \vec{k}_{\rm ph}$ is co-moving wavevector.  We assumed 
$\phi = - \psi$ which is almost always the case \cite{Dodelson} and 
certainly valid during the matter dominated era.  Eq.~(\ref{deneq}) 
shows that the axion BEC has Jeans length
\begin{eqnarray}
k_{\rm J}^{-1} &=& (16 \pi G \rho m^2)^{-{1 \over 4}}\nonumber\\
&=& 1.02 \cdot 10^{14}~{\rm cm}
\left({10^{-5}~{\rm eV} \over m}\right)^{1 \over 2}
\left({10^{-29}~{\rm g/cm^3} \over \rho}\right)^{1 \over 4}~\ .
\label{Jeans}
\end{eqnarray}
The Jeans length is small compared to the smallest scales ($\sim$ 100 kpc) 
for which we have observations on the behavior of density perturbations in 
the linear regime.  Thus axion BEC and CDM are indistinguishable in arena 2
on all scales of observational interest.

In the non-linear regime of structure formation (arena 3) and 
in the absence of rethermalization, the relevant equations are 
\begin{eqnarray}
\partial_t \rho + \vec\nabla \cdot (\rho \vec{v}) = 0~&,&~
\vec\nabla \times \vec{v} = 0\nonumber\\
\partial_t \vec{v} + (\vec{v} \cdot \vec\nabla) \vec{v} &=& 
- \vec\nabla \psi - \vec\nabla q~~~\ . 
\label{nonlin}
\end{eqnarray}
Eqs.~(\ref{nonlin}) are equivalent to the Schr\"{o}dinger equation 
for particles in a Newtonian gravitational field.  Axion BEC and CDM 
differ in that the $-\vec{\nabla} q$ term is absent from the force 
law for CDM.  However, as was shown by numerical simulation \cite{WK}, 
and as is expected from the WKB approximation, the differences occur 
only on length scales smaller than the de Broglie wavelength.   Since 
the axion de Broglie wavelength (of order 10 meters in galactic halos) 
is negligbly small compared to all length scales of observational 
interest, we again find that axion BEC and CDM are indistinguishable 
when there is no rethermalization of the BEC.

However, we found that gravitational interactions do rethermalize the 
axion BEC continually so that the axion state tracks the lowest energy 
state.  This is relevant to the angular momentum distribution of dark 
matter axions in galactic halos.  The angular momentum of galaxies is 
caused by the gravitational torque of nearby galaxies early on when 
protogalaxies are still close to one another \cite{Peebles}.  The 
angular momentum distribution acquired by the dark matter particles 
determines the structure of the inner caustics that they form in 
galactic halos \cite{sing,inner}.  If that distribution is characterized 
by net overall rotation, implying $\vec{\nabla} \times \vec{v} \neq 0$, 
the inner caustics are a set of ``tricusp rings" \cite{sing}.  If the 
velocity field is irrotational ($\vec{\nabla} \times \vec{v} = 0$), 
the inner caustics have a tent-like structure \cite{inner} quite 
distinct from that of tricusp rings.  Evidence has been found for 
tricusp rings \cite{MWhalo}, as opposed to the tent-like caustics
of the $\vec{\nabla} \times \vec{v} = 0$ case.  This raises a
puzzle for CDM.  Indeed one can show \cite{inner} that the velocity 
field of ordinary cold dark matter, such as WIMPs, remains irrotational 
as it is the result of gravitational forces which are proportional to 
the gradient of the Newtonian potential. The puzzle is solved if the 
dark matter is an axion BEC which rethermalizes while tidal torque 
is applied to it.  Indeed, the lowest energy state for given total 
angular momentum is one in which each particle carries an equal 
amount of angular momentum.  In that case there is net overall 
rotation.  $\vec{\nabla} \times \vec{v} \neq 0$ is accomodated 
in the BEC through the appearance of vortices.  The phenomenon 
is observed in quantum liquids and well understood \cite{Pethik}.

Finally we consider the behaviour of density perturbations as they enter 
the horizon (arena 1).  Here too axion BEC may differ from CDM.  The CDM 
perturbations evolve linearly at all times.  The axion BEC perturbations 
do not evolve linearly when they enter the horizon because the condensates 
which prevailed in neighboring horizon volumes rearrange themselves, through 
their gravitational interactions, into a new condensate for the expanded 
horizon volume.  This produces local correlations between modes of different 
wavevector since the perturbation of wavevector $\vec{k}$, upon entering 
the horizon, is determined by the perturbations of wavevector say 
${1 \over 2}\vec{k}$ in its neighborhood.  We propose this as a 
mechanism for the alignment of CMBR anisotropy multipoles \cite{align} 
through the integrated Sachs-Wolfe (ISW) effect. Unlike CDM, the ISW 
effect is large in axion BEC because the Newtonian potential $\psi$
changes entirely after entering the horizon in response to the 
rearrangement of the axion BEC.

We conclude that a case can be made that a large fraction of the dark
matter is axions.  Although the QCD axion is best motivated, a large
class of axion-like particles has the properties described here.

We thank Georg Raffelt, Larry Widrow and Edward Witten for alerting us 
to errors in earlier versions of our paper, and for stimulating comments.  
This work was supported in part by the U.S. Department of Energy under 
contract DE-FG02-97ER41029.


\begin{thebibliography}{}

\bibitem{dmBEC1}
S.-J. Sin, Phys. Rev. D50 (1994) 3650; 
J. Goodman, New Astronomy Reviews 5 (2000) 103; 
W. Hu, R. Barkana and A. Gruzinov, Phys. Rev. Lett. 85 (2000) 1158;
J.-W. Lee and S. Lim, arXiv:0812.1342 and references therein;
E.W. Mielke and J.A. V\'elez P\'erez, Phys. Lett. B671 (2009) 174.

\bibitem{dmBEC2}
F. Ferrer and J.A. Grifols, JCAP 12 (2004) 012;
C.G. Bohmer and T. Harko, JCAP 06 (2007) 025.

\bibitem{axion}
R. D. Peccei and H. Quinn, Phys. Rev. Lett. {\bf 38} (1977) 1440 and Phys.
Rev. {\bf D16} (1977) 1791; S. Weinberg, Phys. Rev. Lett. {\bf 40}
(1978) 223; F. Wilczek, Phys. Rev. Lett. {\bf 40} (1978) 279.

\bibitem{invis}
J. Kim, Phys. Rev. Lett. {\bf 43} (1979) 103; M. A. Shifman, 
A. I. Vainshtein and V. I. Zakharov, Nucl. Phys. {\bf B166} (1980) 493; 
A. P. Zhitnitskii, Sov. J. Nucl. {\bf 31} (1980) 260;  M. Dine, 
W. Fischler and M. Srednicki, Phys. Lett. {\bf B104} (1981) 199.

\bibitem{axrev}
J.E. Kim, Phys. Rep. {\bf 150} (1987) 1;
M.S. Turner, Phys. Rep. {\bf 197} (1990) 67;
G.G. Raffelt, Phys. Rep. {\bf 198} (1990) 1.

\bibitem{axdm}
J. Preskill, M. Wise and F. Wilczek, Phys. Lett. {\bf B120} (1983) 127;
L. Abbott and P. Sikivie, Phys. Lett. {\bf B120} (1983) 133;
M. Dine and W. Fischler, Phys. Lett. {\bf B120} (1983) 137.

\bibitem{axcos}
P. Sikivie, Lect. Notes Phys. 741 (2008) 19.

\bibitem{ST}
D.V. Semikoz and I.I. Tkachev, Phys. Rev. Lett. 74 (1995) 3093 and 
Phys. Rev. D55 (1997) 489.  See also: S. Khlebnikov, Phys. Rev. A66
(2002) 063606 and references therein.

\bibitem{sound}
N.N. Bogoliubov, J. Phys. (Moscow) 11 (1947) 23; 
T.D. Lee, K. Huang and C.N. Yang, Phys. Rev. 106 (1957) 1135;
J. Bernstein and S. Dodelson, Phys. Rev. Lett. 66 (1991) 683.

\bibitem{Pethik}
C.J. Pethik and H. Smith, {\it Bose-Einstein Condensation in 
Dilute Gases}, Cambridge University Press 2002.

\bibitem{Dodelson}
S. Dodelson, {\it Modern Cosmology}, Academic Press 2003.

\bibitem{WK}
L.M. Widrow and N. Kaiser, Ap. J. 416 (1993) L71.

\bibitem{Peebles}
P.J.E. Peebles, Ap. J. 155 (1969) 393.

\bibitem{sing}
P. Sikivie, Phys. Rev. D60 (1999) 063501.

\bibitem{inner}
A. Natarajan and P. Sikivie, Phys. Rev. D73 (2006) 023510.

\bibitem{MWhalo}
L.D. Duffy and P. Sikivie, Phys. Rev. D78 (2008) 063508, 
and references therein.

\bibitem{align}
M. Tegmark, A. de Oliveira-Costa and A. Hamilton, Phys. Rev. D68 (2003) 123523;
A. de Oliveira-Costa, M. Tegmark, M. Zaldarriaga and A. Hamilton, Phys. Rev.
D69 (2004) 063516; C.J. Copi, D. Huterer, D.J. Schwarz and G.D. Starkman, 
MNRAS 367 (2006) 79 and references therein.

\end{thebibliography}
\end{document}